\documentclass[a4paper,11pt]{article}
\usepackage{pos}
\RequirePackage{xspace}
\RequirePackage{relsize}
\usepackage{booktabs}
\usepackage{listings}

\title{Run Dependent Monte Carlo at Belle~II}

\author*[a,b]{Giovanni Gaudino}

\affiliation[a]{SSM,\\
  Via Mezzocannone, 4, 8018, Napoli, Italia}

\affiliation[b]{INFN,\\
via Cintia, 80126, Napoli, Italia}

\emailAdd{gaudino@na.infn.it}

\abstract{The Belle II experiment at the SuperKEKB accelerator in Tsukuba, Japan, searches for physics beyond the Standard Model, with a focus on precise measurements of flavor physics observables. Highly accurate Monte Carlo simulations are essential for this endeavor, as they must correctly model the variations in detector conditions and beam backgrounds that occur during data collection. To meet this requirement, the "run-dependent" Monte Carlo has been developed. This approach incorporates time-dependent detector conditions and beam-induced backgrounds collected via random triggers, allowing for different conditions with a granularity of just a few hours. In this document, we will discuss the procedures and challenges associated with producing run-dependent Monte Carlo simulations for Belle II. We will also highlight the improvements these simulations offer over traditional "run-independent" Monte Carlo methods.}

\FullConference{The European Physical Society Conference on High Energy Physics (EPS-HEP2025)\\
7-11 July 2025\\
Marseille, France\\}

\def\belletwo{\mbox{Belle~II}\xspace}
\def\invcm  {\ensuremath{{\text \,cm}}\xspace}
\def\nb {\ensuremath{\text \,nb}\xspace}
\def\invfb   {\ensuremath{\mbox{\,fb}^{-1}}\xspace}
\def\invab   {\ensuremath{\mbox{\,ab}^{-1}}\xspace}
\def\hz   {\ensuremath{{\text \,Hz}}\xspace}
\def\invsec  {\ensuremath{{\text \,s}^{-1}}\xspace}

\def\Pgamma      {\ensuremath{\gamma}\xspace}                 
\def\Pe          {\ensuremath{e}\xspace}
\def\Pmu         {\ensuremath{\mu}\xspace}                 
\def\Ptau        {\ensuremath{\tau}\xspace}                 

\def\epem       {\ensuremath{\Pe^+\Pe^-}\xspace}
\def\mumu       {\ensuremath{\Pmu^+\Pmu^-}\xspace}
\def\tautau     {\ensuremath{\Ptau^+\Ptau^-}\xspace}

\def\Pq         {\ensuremath{q}\xspace}
\def\qqbar      {\ensuremath{\Pq\overline{\Pq}}\xspace}
\def\BBbar      {\ensuremath{B\overline{B}}\xspace}

\def\basf{\texttt{basf2}\xspace}

\begin{document}
\maketitle

\section{Introduction}
The Belle~II experiment, located at the SuperKEKB accelerator in Tsukuba, Japan, is a high-energy particle physics experiment designed to explore the fundamental properties of matter and antimatter. It is the successor to the original Belle experiment, which operated from 1999 to 2010 and made significant contributions to the field of flavor physics, particularly in the study of $B$ meson decays. The Belle~II experiment aims to build upon these achievements by collecting a much larger dataset, with the goal of achieving a tenfold increase in statistics compared to its predecessor. This ambitious goal is made possible by the SuperKEKB accelerator, which is designed to deliver a peak luminosity of $6.4\times 10^{35}\invcm\invsec$, significantly higher than the $2\times 10^{34}\invcm\invsec$ achieved by the original Belle experiment~\cite{Akai:2018mbz, Abe:2010gxa}.

The Belle~II detector is a sophisticated particle physics apparatus that includes a variety of sub-detectors, each designed to measure different aspects of the particles produced in high-energy collisions. The detector is equipped with a silicon vertex detector, a drift chamber, an electromagnetic calorimeter, and a muon system, among other components. These detectors work together to reconstruct the trajectories and properties of particles produced in collisions, allowing physicists to study rare decays and search for new physics beyond the Standard Model~\cite{Kou:2018nap}.

To achieve its scientific goals, the Belle~II experiment relies heavily on Monte Carlo (MC) simulations. These simulations are essential for understanding the detector response, modeling the background processes, and calibrating the data collected during experiments. The Belle~II collaboration has developed a sophisticated MC production framework that allows for the generation of realistic simulated events, which are then used to compare with the experimental data.

In this paper, we will focus on the run-dependent Monte Carlo (MCrd) production at Belle~II, which is a crucial component of the experiment's data analysis and calibration efforts. The MCrd approach is designed to account for the time-dependent variations in detector conditions and beam backgrounds that occur during data collection. This is particularly important for high-precision measurements, where even small discrepancies between data and simulation can lead to significant systematic uncertainties. By incorporating run-specific conditions and beam backgrounds, the MCrd samples provide a more accurate representation of the experimental environment.\\
The paper is organized as follows: in Section~\ref{sec:workflow}, we will describe the data production workflow at Belle~II, including the steps involved in transferring raw data, reconstructing events, and producing analysis-level datasets. In Section~\ref{sec:mcrd}, we will delve into the specifics of the MCrd production process, including the categorization of physics channels, background preparation, detector configuration retrieval, and job submission to the distributed computing infrastructure. Finally, in Section~\ref{sec:summary}, we will summarize the key points discussed in this paper and highlight the significance of the MCrd approach for the Belle~II experiment.

\section{Data Production Workflow}
\label{sec:workflow}
\begin{figure}[t]
    \centering
    \includegraphics[width=0.8\textwidth]{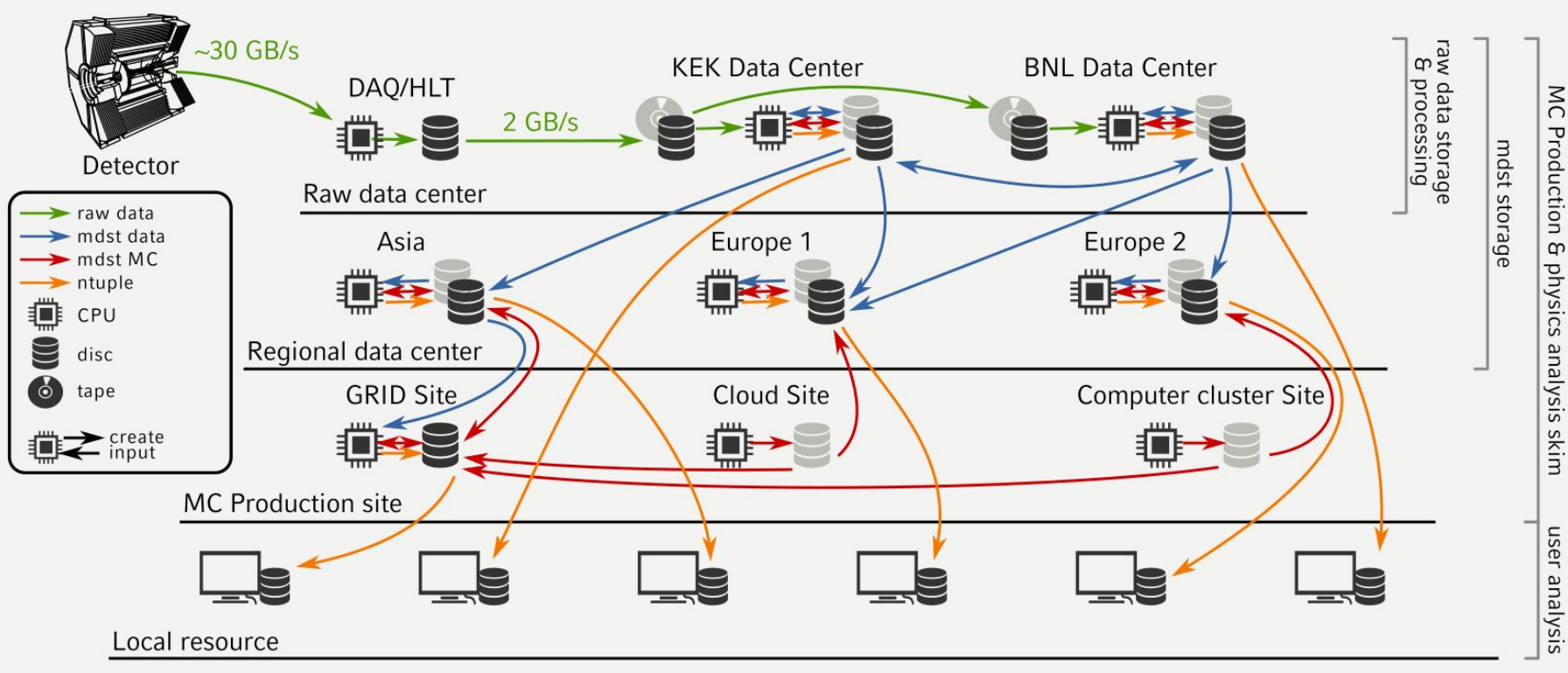}
    \caption{Schematic overview of the \belletwo Grid computing infrastructure, illustrating the data flow from raw acquisition through reconstruction to analysis-ready datasets.}
    \label{fig:grid_schema}
\end{figure}

The \belletwo data production workflow encompasses a comprehensive multi-stage pipeline engineered to ensure high-fidelity reconstruction and precise calibration of experimental data. The workflow architecture is illustrated schematically in Fig.~\ref{fig:grid_schema}.

\subsection{Data Transfer and Grid Registration}
Raw data acquisition begins with the transfer of collision events from the online data acquisition system to the KEK Central Computing system (KEKCC) at the KEK laboratory in Tsukuba, Japan. Following archival at KEKCC, the \belletwo core computing team registers datasets within the distributed computing infrastructure managed by the \belletwo Grid.

\subsection{Reconstruction and Calibration Streams}
The reconstruction process initiates with High-Level Trigger (HLT) skimmed samples, which serve as the foundation for detector calibration procedures. Among these, specialized calibration streams are generated, including "delayed Bhabha" samples—events recorded at approximately $5\hz$ following Bhabha scattering interactions. The reduced acquisition rate accommodates the computational overhead associated with preserving complete electromagnetic calorimeter (ECL) waveform data.

Calibration proceeds through a two-tier approach: \emph{prompt calibration} provides rapid, expert-validated detector configuration parameters, while subsequent \emph{full calibration} allows to fix issues present in the prompt calibration and to add calibrations developed after the data taking. The resulting calibration constants, termed \emph{payloads}, are systematically stored in the \belletwo conditions database, enabling run-specific detector modeling.

\subsection{Monte Carlo Production Framework}
The \belletwo collaboration maintains an annual Monte Carlo production schedule synchronized with software release cycles. Two complementary simulation approaches are employed:

\paragraph{Run-Independent Monte Carlo (MCri):} These simulations utilize time-averaged detector configurations and idealized background models derived from dedicated studies. While computationally efficient and suitable for general analyses, MCri samples may exhibit data-simulation discrepancies in precision measurements due to their averaged nature. Current MCri production reaches approximately $3\invab$ of integrated luminosity.

\paragraph{Run-Dependent Monte Carlo (MCrd):} In contrast, MCrd samples incorporate run-specific detector configurations and data-driven beam background modeling, providing enhanced fidelity for systematic uncertainty control. Background contributions are derived from real-time sub-detector measurements, while detector geometry and alignment precisely match experimental conditions during specific data-taking periods. The current MCrd dataset corresponds to approximately four times the recorded data luminosity, totaling $\sim2\invab$.

The computational demands of MCrd production are substantially higher than MCri due to the inclusion of detailed run-dependent information and realistic background overlay procedures. However, this investment enables high-precision physics measurements that fully exploit the \belletwo dataset capabilities.

\section{Run-Dependent Monte Carlo Production}
\label{sec:mcrd}

The production of run-dependent Monte Carlo (MCrd) samples represents a significantly more complex and computationally intensive process compared to conventional run-independent simulations. This complexity stems from the requirement to accurately reproduce the time-varying detector and accelerator conditions that characterize each data collection period. The \belletwo experiment organizes data acquisition through a hierarchical structure of \emph{experiment} and \emph{run} numbers, where experiment numbers denote major configuration changes (detector upgrades, beam condition modifications, trigger setting adjustments) and run numbers represent the fundamental units of consistent data-taking conditions.

The MCrd production framework mirrors this organizational structure and encompasses several critical stages: physics channel categorization, background overlay preparation, detector configuration retrieval, and distributed job submission to the \belletwo Grid infrastructure.

\subsection{Physics Channel Classification}

MCrd samples are systematically categorized into two primary classes based on their intended application:

\paragraph{Generic Samples} encompass fundamental Standard Model processes including continuum quark-antiquark production (\epem$\to$\qqbar), tau pair production (\tautau), $B$-meson pair production (\BBbar), low-multiplicity events (fewer than five reconstructed tracks), and electromagnetic processes such as dimuon and Bhabha scattering~\cite{Jadach:1990mz, Jadach:1999vf, Lange:2001uf, Agostinelli:2002hh, davidson2015photos, BARBERIO1991115, Sjostrand:2014zea}. Production volumes for these channels are determined by target integrated luminosity requirements, with scaling factors applied to achieve optimal statistical precision for systematic studies.

Table~\ref{tab:generic_samples} summarizes the current production campaign parameters, including cross sections and luminosity scaling factors for each generic channel.

\begin{table}[t]
    \centering
    \caption{Production parameters for generic MCrd samples at \belletwo. Cross sections ($\sigma$) are given in nanobarns, with MC-to-data scaling factors indicating the statistical enhancement relative to recorded data. The $\Pgamma\Pgamma$ process represents diphoton production; \texttt{llXX} and \texttt{hhISR} denote multi-lepton and hadronic initial-state radiation final states, respectively.}
    \label{tab:generic_samples}
    \begin{tabular}{lcc}
        \toprule
        \toprule
        Channel & Cross section $\sigma$ [\nb] & MC/data factor \\
        \midrule
        \qqbar     & 3.68   & 4 \\
        \tautau    & 0.919  & 4 \\
        \BBbar     & 1.05   & 4 \\
        \mumu      & 1.148  & 1 \\
        \epem      & 295.8  & 0.1 \\
        \epem\epem & 39.55  & 1 \\
        \epem\mumu & 18.83  & 1 \\
        \Pgamma\Pgamma & 5.10 & 2 \\
        \texttt{llXX}  & 2.01 & 4 \\
        \texttt{hhISR} & 0.216 & 1 \\
        \bottomrule
        \bottomrule
    \end{tabular}
\end{table}

\paragraph{Signal Samples} correspond to specific decay channels targeted by dedicated physics analyses. Unlike generic samples, signal production is governed by fixed event statistics rather than integrated luminosity constraints. While the total number of signal requests substantially exceeds that of generic channels (approximately 200 active requests), the reduced event statistics per sample ensures computational feasibility.

\subsection{Background Modeling and Overlay}

Realistic beam-induced background incorporation constitutes a cornerstone of MCrd fidelity. Background characterization employs a data-driven methodology utilizing specialized trigger configurations activated with predetermined delays following Bhabha scattering events. This temporal separation ensures isolation of pure background contributions uncontaminated by primary collision products.

Background-enriched events are acquired at reduced rates and subsequently processed to extract raw digitized detector responses from tracking and calorimetry subsystems. During simulation, this digit-level background information is overlaid onto Monte Carlo-generated signal events, reproducing realistic detector occupancy and noise conditions characteristic of actual data collection periods.

Background preparation is performed independently for each experiment period. While computationally modest compared to full event simulation, this process is essential for accurate reproduction of the time-dependent operational environment of the \belletwo detector.

\subsection{Detector Configuration Management}

Detector configuration and calibration constant preparation represents the most resource-intensive component of the MCrd production pipeline. Each \belletwo subsystem must provide run-specific calibration payloads encompassing per example alignment parameters, dead channel mappings, and time-dependent gain calibrations.

These payloads exhibit varying temporal stability: dead channel maps remain relatively stable throughout extended data-taking periods, while collision point coordinates and beam energy calibrations require frequent updates reflecting evolving accelerator conditions.

The aggregation, validation, and distribution of conditions data constitutes the primary bottleneck in the production workflow. Ongoing optimization efforts focus on streamlining payload management and validation procedures for future production campaigns.

\subsection{Grid-Based Production Infrastructure}

The final production stage leverages the distributed computing capabilities of the \belletwo Grid infrastructure for large-scale event generation. Job submission utilizes structured JSON configuration files specifying event requirements for individual runs, with luminosity-weighted event distributions ensuring statistical representativity.

Each run necessitates independent simulation jobs due to unique detector and background conditions, resulting in substantial computational loads distributed across Grid resources. The simulation framework employs the \basf software package, providing comprehensive detector modeling and physics event generation capabilities.

Prior to full-scale production deployment, extensive validation testing is conducted on dedicated computing infrastructure to verify job stability and output quality. Following successful validation, jobs are distributed to Grid computing centers, with resulting MCrd samples systematically catalogued and made available for physics analysis workflows.

\section{Data Management and Book-keeping}

Systematic organization and management of MCrd samples represents a critical infrastructure component ensuring efficient data access and reproducible analyses across the \belletwo collaboration.

\subsection{Collection-Based Organization}

The \belletwo Monte Carlo framework employs a hierarchical \emph{collection} system for dataset organization. Collections aggregate related Monte Carlo samples corresponding to specific physics processes within well-defined experimental periods. When submitting computational jobs to the \belletwo Grid infrastructure, analysts reference these collections to specify input datasets, ensuring consistency and reproducibility across diverse analysis workflows.

Each collection encompasses all Monte Carlo datasets for a particular physics channel within a designated data-taking period. The hierarchical naming convention follows a structured metadata encoding scheme, for example

\begin{center}
\begin{minipage}{0.8\textwidth}
\begin{lstlisting}[language=bash, basicstyle=\ttfamily\footnotesize, backgroundcolor=\color{gray!10}, frame=single]
/belle/collection/MC/MC16rd_proc16_ddbar_Run1_4S_v1
\end{lstlisting}
\end{minipage}
\end{center}

This systematic nomenclature encodes essential parameters including the Monte Carlo campaign identifier (\texttt{MC16rd}), processing version (\texttt{proc16}), physics channel (\texttt{ddbar}), data-taking period (\texttt{Run1}), collision energy configuration (\texttt{4S}), and dataset version (\texttt{v1}).

\subsection{Metadata Schema}

Each collection maintains comprehensive metadata providing analysts with essential information for proper dataset utilization. Current metadata implementations include fundamental parameters such as integrated luminosity and data processing levels:

\begin{center}
\begin{minipage}{0.8\textwidth}
\begin{lstlisting}[language=bash, basicstyle=\ttfamily\footnotesize, backgroundcolor=\color{gray!10}, frame=single]
##########Metadata of Collection###############
name: /belle/collection/MC/MC16rd_proc16_ddbar_Run1_4S_v1
dataType: mc
dataLevel: mdst
description: Collection for MC16rd_proc16-Run1-ddbar-4S
int_luminosity: 1429.23/fb
#################################################
\end{lstlisting}
\end{minipage}
\end{center}

Future metadata schema enhancements will incorporate additional parameters including event statistics, cross sections, and detailed production configuration information to further facilitate analysis planning and systematic uncertainty evaluation.

\section{Summary}
\label{sec:summary}

The \belletwo experiment is acquiring an unprecedented volume of data, making the accurate and efficient production of Monte Carlo samples a critical component of the collaboration’s physics program. To meet the stringent requirements of high-precision measurements, a dedicated run-dependent Monte Carlo production framework has been developed. Although the system is inherently complex and computationally demanding, it provides the most realistic simulation environment currently available within the collaboration.

By incorporating run-by-run variations in detector conditions and beam backgrounds, the MCrd samples enable a significant reduction in discrepancies between data and simulation. This improved fidelity is essential for controlling systematic uncertainties and enhancing the reliability of physics results.

As of today, \belletwo has recorded approximately $500\invfb$ of data, while the corresponding MCrd samples produced for high-multiplicity processes already amount to an integrated luminosity of about $1700\invfb$, for the high multiplicity samples. This substantial MC production effort underscores the collaboration’s commitment to high-quality simulation and robust data analysis, laying a strong foundation for future discoveries in flavor physics.\\
A detailed overview on the Monte Carlo run dependent production at \belletwo was presented at 2025 EPS-HEP conference~\cite{mcrd_eps}.


\begin{thebibliography}{99}
\bibitem{mcrd_eps}
Giovanni Gaudino for the Belle~II Data Production group,
\textit{Run Dependent Monte Carlo at Belle~II},
Contribution to EPS-HEP 2025, Marseille, France (2025).
\url{https://indico.in2p3.fr/event/33627/contributions/155250/}



\bibitem{Abe:2010gxa}
T.~Abe \textit{et al.} [Belle~II collaboration],
``Belle II Technical Design Report,''
arXiv:1011.0352 [physics.ins-det].

\bibitem{Akai:2018mbz}
K.~Akai, K.~Furukawa and H.~Koiso,
``SuperKEKB collider,''
Nucl.\ Instrum.\ Meth.\ A \textbf{907}, 188 (2018),
doi:10.1016/j.nima.2018.08.017.

\bibitem{Kou:2018nap}
W.~Altmannshofer \textit{et al.}, E.~Kou and P.~Urquijo, eds.,
``The Belle~II physics book,''
PTEP \textbf{2019}, no.~12, 123C01 (2019),
Erratum: PTEP \textbf{2020}, 029201 (2020),
doi:10.1093/ptep/ptz106, 10.1093/ptep/ptaa008,
arXiv:1808.10567 [hep-ex].





\bibitem{Jadach:1990mz}
S.~Jadach, J.~H.~Kuhn and Z.~W\c{a}s,
``TAUOLA: A library of Monte Carlo programs to simulate decays of polarized tau leptons,''
Comput.\ Phys.\ Commun.\ \textbf{64}, 275 (1990),
doi:10.1016/0010-4655(91)90038-M.

\bibitem{Jadach:1999vf}
S.~Jadach, B.~F.~L.~Ward and Z.~W\c{a}s,
``The precision Monte Carlo event generator KK for two-fermion final states in $e^+e^-$ collisions,''
Comput.\ Phys.\ Commun.\ \textbf{130}, 260 (2000),
doi:10.1016/S0010-4655(00)00048-5,
arXiv:hep-ph/9912214.

\bibitem{Lange:2001uf}
D.~J.~Lange,
``The EvtGen particle decay simulation package,''
Nucl.\ Instrum.\ Meth.\ A \textbf{462}, 152 (2001),
doi:10.1016/S0168-9002(01)00089-4.

\bibitem{Agostinelli:2002hh}
S.~Agostinelli \textit{et al.} [GEANT4 collaboration],
``GEANT4: A simulation toolkit,''
Nucl.\ Instrum.\ Meth.\ A \textbf{506}, 250 (2003),
doi:10.1016/S0168-9002(03)01368-8.



\bibitem{davidson2015photos}
N.~Davidson, T.~Przedzinski and Z.~Was,
``PHOTOS Interface in C++; Technical and Physics Documentation,''
arXiv:1011.0937 [hep-ph].

\bibitem{BARBERIO1991115}
E.~Barberio, B.~van~Eijk and Z.~Was,
``Photos — a universal Monte Carlo for QED radiative corrections in decays,''
Comput.\ Phys.\ Commun.\ \textbf{66}, 115 (1991),
doi:10.1016/0010-4655(91)90012-A.

\bibitem{Sjostrand:2014zea}
T.~Sj\"{o}strand \textit{et al.},
``An Introduction to PYTHIA 8.2,''
Comput.\ Phys.\ Commun.\ \textbf{191}, 159 (2015),
doi:10.1016/j.cpc.2015.01.024,
arXiv:1410.3012 [hep-ph].

\end{thebibliography}
\end{document}